\documentclass[aps,preprint,prd,epsfig]{revtex4}
\usepackage{amssymb}
\usepackage{amsmath}
\usepackage{graphicx}
\usepackage{fancyhdr}
\newcommand{\be}{\begin{equation}}
\newcommand{\ee}{\end{equation}}
\newcommand{\bea}{\begin{eqnarray}}
\newcommand{\eea}{\end{eqnarray}}
\newcommand{\bes}{\begin{subequations}}
\newcommand{\ees}{\end{subequations}}

\newcommand{\bc}{\begin{center}}
\newcommand{\ec}{\end{center}}
\usepackage{amsmath, amssymb} %permite o uso de alguns símbolos matemáticos
\usepackage{slashed}

\begin{document}

\title{Vacuum stability and spontaneous violation of the lepton number at low energy scale in a model for light sterile neutrinos}

\author{Jo\~ao Paulo Pinheiro   and C. A. de S. Pires}
%\email{cpires@fisica.ufpb.br}
\affiliation{{ Departamento de
F\'{\i}sica, Universidade Federal da Para\'\i ba, Caixa Postal 5008, 58051-970,
Jo\~ao Pessoa, PB, Brasil}}

\date{\today}

\begin{abstract}
It is well known that the Standard Model of the Electroweak interactions  rests on a metastable vacuum. This can only be fixed by means of new physics.  Presently neutrino physics provides the most intriguing framework to formulate new physics. This is so because, in addition to the problem  of the lightness of the active standard neutrinos,  currently MiniBooNE experimental result may be indicating that  sterile neutrinos exist and are light, too. In this case, it is reasonable to expect that the framework that yields  light active and sterile neutrinos could stabilize the vacuum, too. In order to achieve this goal, we consider an extension of the standard model which involves new fermions in the form of right-handed neutrinos ($\nu_R$) and new scalars in the form of triplet ($\Delta$) and singlet ($\sigma$). Within this framework, tiny masses are obtained when we consider that lepton number is spontaneously broken at low energy scale which means that $\Delta$ and $\sigma$, both,  develop  very small vacuum expectation values.  We investigate if this setting leads to a stable vacuum. For this we obtain the whole set of conditions over the Quartic Terms of the Potential that ensures that the model  is Bounded From Below(BFB) and  evaluate the RGE-evolution of the self coupling  of the Higgs. We show that in such a scenario  the  Quartic Coupling $\Phi^T \Delta \Phi \sigma$ , where $\Phi$ is the standard Higgs doublet,  is responsible for the stability of the Electroweak Vacuum up to Planck scale. We also extract constraints over the parameters of the Potential by means of Lepton Flavor Violating(LFV) processes and from invisible decay of the standard-like Higgs.
\end{abstract}

\maketitle

\section{Introduction}
%%%
Although people has devoted considerable attention to the study  of  extensions of the standard Higgs sector\cite{2hdm},\cite{cheng},\cite{singlet}, relative few attention has been given to a Higgs sector involving triplet ($\Delta$), doublet ($\Phi$) and singlet ($\sigma$)  of scalars\cite{Schechter},\cite{valle},\cite{cogollo}. From now on we refer to this case as 3-2-1 model. This model is interesting by its own right. However it get even more interesting when right-handed neutrinos are introduced, too. This is so because in this case the 3-2-1 model yields the most general neutrino mass matrix involving Majorana and Dirac mass terms for both neutrinos. Hence, when we assume that  lepton number is spontaneously violated at low energy scale, right-handed neutrinos acquire light masses and may  explain the recent  MiniBooNE experimental result\cite{miniboone}  by means of neutrino oscillation. 

In this work we derive the complete set of conditions that guarantee the Potential of the 3-2-1 model to be BFB. For the specific case when lepton number is spontaneously broken at low energy scale, we obtain the spectrum of scalars of the model and discuss the stability of the vacuum  by evaluating the RGE- evolution of the self-coupling of the standard-like Higgs up to Planck scales . This case is particularly interesting because it encompasses  a Majoron and a light CP-even scalar in their spectrum of scalars. We discuss the contributions of these scalars for the  invisible decay channels of the standard-like Higgs and of the neutral gauge boson $Z$.  We also obtain the constraints that LFV put over the parameters of the Potential. In what concern neutrino physics, we provide a solution, i.e., a set of values for the Yukawa couplings, that recovers the standard neutrino sector and provides at least one right-handed neutrino with mass resting on eV scale and robustly mixed with  the standard neutrinos in such a way that accommodates MiniBooNE current results by means of neutrino oscillation and is in agreement with cosmological data.

This work is organized as follows. In Sec. II we develop the main aspect of the model including neutrino masses, while in Sec. III we develop the scalar sector. In Sec. IV we discuss the stability of the vacuum. In Sec. V we present our final remarks.

\section{The 3-2-1 model}
The  leptonic sector of the model is composed by the standard doublet $L$ plus right-handed neutrinos in the singlet form,
\begin{equation}
    L_{i}=
\left (
\begin{array}{c}
\nu_i \\
\ell_i
\end{array}
\right )_L \,\, ;\,\,\,\, \ell_{i_R} ;\,\,\,\, \nu_{i_R}; 
\label{lept-lang} 
\end{equation}
where $i=e\,,\mu\,,\,\tau$, while the standard scalar sector is composed by one triplet, one doublet and one singlet of scalars,
\begin{equation}
\Delta=
\begin{pmatrix}
\Delta^{0} & \dfrac{\Delta^{+}}{\sqrt{2}}\\
\dfrac{\Delta^{+}}{\sqrt{2}}& \Delta^{++}
\end{pmatrix};\; \; \;
\Phi=
\begin{pmatrix}
\phi^{0} \\
\phi^{-}
\end{pmatrix}; \; \; \;
\sigma.
\end{equation}

The quark sector is the standard one.

The most general potential involving this scalar content and that conserves lepton number is composed by the following terms
\begin{eqnarray}
&&V(\sigma, \Phi, \Delta) = \mu_1 ^{2} \sigma^{*}\sigma + 	\mu_2^{2} \Phi^{\dagger}\Phi + \mu_3^{2} tr(\Delta^{\dagger}\Delta) \nonumber \\
&&+\lambda_1 (\Phi^{\dagger}\Phi)^{2} + \lambda_2 [tr(\Delta^{\dagger} \Delta)]^2\nonumber 
+ \lambda_3\Phi^{\dagger}\Phi tr(\Delta^{\dagger} \Delta)\nonumber \\ 
&&+ \lambda_4 tr(\Delta^{\dagger}\Delta\Delta^{\dagger} \Delta) + \lambda_5 (\Phi^{\dagger}\Delta^{\dagger}\Delta\Phi)+	 \nonumber\\
&&\beta_1 (\sigma^{*}\sigma)^2 + \beta_2\Phi^{\dagger}\Phi\sigma^{*}\sigma + \beta_3 tr(\Delta^{\dagger} \Delta)
\sigma^{*}\sigma- \nonumber \\ &&\kappa(\Phi^{T}\Delta\Phi\sigma + H.c.).
\label{potential}
\end{eqnarray}

With such lepton and scalar content, the Yukawa interactions  that generate mass for all neutrinos of the model is given by

\begin{eqnarray}
{\cal L}^\nu_Y =\frac{1}{2} Y^{L}_{ij}L_{i}^{T}\Delta L_{j} + Y^{D}_{ij}\overline{L_{i}}\Tilde{\phi}\nu_{Rj} + \frac{1}{2}Y^{R}_{ij}\overline{\nu^{C}_{Ri}} \nu_{Rj}\sigma  + H.c..
\label{GYI}
\end{eqnarray}
The Yukawa interactions of the charged fermions are the standard ones.

When the neutral scalars of the model develop vacuum expectation values (VEV) different from zero, i.e, $\langle \sigma\rangle=\frac{v_1}{\sqrt{2}}$, $\langle \phi\rangle=\frac{v_2}{\sqrt{2}}$ and $\langle \Delta\rangle=\frac{v_3}{\sqrt{2}}$, the Yukawa interactions in Eq. (\ref{GYI}) provide the following mass terms for the neutrinos,

\begin{eqnarray}
{\cal L}^{D+M}_{mass} = \frac{1}{2}\bar \nu_L^C M_L\nu_L + \bar \nu_L M_D \nu_R + \frac{1}{2}\bar \nu_R ^C M_R \nu_R  + H.c.,
\label{NMT}
\end{eqnarray}
with $\nu_L =(\nu_{e_L}\,\,\, \nu_{\mu_L}\,\,\, \nu_{\tau_L}) ^T$ and $\nu_R =(\nu_{e_R}\,\,\, \nu_{\mu_R}\,\,\, \nu_{\tau_R}) ^T$.

Considering the basis $\nu=(\nu_L\,\,\,\,\,\nu_R^C  )^T$,  we can simplify Eq. (\ref{NMT}) to 
\begin{eqnarray}
{\cal L}^{D+M}_{mass} = \frac{1}{2} \bar \nu^C M^{D+M} \nu + H.c.,
\end{eqnarray}
and the 6$\times$6 symmetric mass matrix is given by 
\begin{equation}
 M^{D+M} = \left(\begin{array}{cc}
M_L & M_{D}^{T}        \\
\newline               \\
M_{D} & M_{R}
\end{array}\right),
\label{M+D}
\end{equation}
where $M_L =Y^L \langle \Delta \rangle$, $M_D= Y^D \langle \phi \rangle$ and $M_R =Y^R \langle \sigma \rangle$. $M^{D+M}$ is the most general neutrino mass matrix. It involves Dirac and Majorana mass terms for both left and right-handed neutrinos. The 3-2-1 model is the simplest model that generates this mass matrix in the case of spontaneous violation of the lepton number. 

The relation among the flavor basis, $\nu $, with the physical ones, $N=(N_1 \,\, N_2 \,\, N_3 \,\, N_4 \,\, N_5 \,\, N_6)^T $, is given by  $N=U\nu $ with $U$ being the unitary matrix that diagonalize $M^{D+M}$,
\begin{equation}
    U^T M^{D+M} U=M=\mbox{diag}(M_1\,\,\, M_2),
    \label{mixing}
\end{equation}
where $M_1=\mbox{diag}(m_1\,\,\,m_2\,\,\,m_3)^T$ and $M_2=\mbox{diag}(m_4\,\,\,m_5\,\,\,m_6)^T$.

In order to go further we need to obtain information about the VEVs $v_1$, $v_2$ and $v_3$.  For this we have to develop the scalar sector of the model.   

Firstly, we expand the neutral scalar fields around their respective VEVs,
\begin{eqnarray}
&&\sigma = \dfrac{v_1}{\sqrt{2}} + \dfrac{R_1 + I_1}{\sqrt{2}} \nonumber \\
&&\phi^{0} = \dfrac{v_2}{\sqrt{2}} + \dfrac{R_2 + I_2}{\sqrt{2}} \nonumber \\
&&\Delta^{0} = \dfrac{v_3}{\sqrt{2}} + \dfrac{R_3 + I_3}{\sqrt{2}},
 \end{eqnarray}
 and obtain the set of minimum conditions required by the potential above to allow spontaneous breaking of the symmetries of the model which include the global  $B-L$ symmetry,
\begin{eqnarray}
&& v_1(\mu_1^{2} + \beta_1 v_1^2 + \dfrac{1}{2}\beta_2 v_2^2 + \dfrac{1}{2}\beta_3 v_3^2) - \dfrac{1}{2} \kappa v_2^2 v_3 = 0 \nonumber \\
&& v_2(\mu_2^{2} + \lambda_1 v_2^2 + \dfrac{1}{2}\lambda_3 v_3^2 + \dfrac{1}{2}\lambda_5 v_3^2 + \dfrac{1}{2}\beta_2 v_1^2  -\kappa v_1 v_3) = 0 \nonumber \\
&& v_3(\mu_3^{2} + \lambda_2 v_3^2 + \dfrac{1}{2}\lambda_3 v_2^2 + \lambda_4 v_3^2 + \dfrac{1}{2}\lambda_5 v_2^ 2+  \dfrac{1}{2}\beta_3 v_1^2 )-\dfrac{\kappa v_1 v_2^2}{2} = 0.
\label{minimumEQS}
 \end{eqnarray}

On analyzing this set of constraint, observe that the first and third relations provide
\begin{equation}
    v_1\approx \frac{\mu_3}{\mu_1}v_3.
    \label{v3xv1}
\end{equation} 
This relation is interesting because it relates $v_3$, which has an upper bound of 2 GeV, with $v_1$ that is free to develop any value. According to this relation, if we assume that  $\mu_3 \sim \mu_1$ than we have $v_1 \sim v_3$.  Any hierarchy among $v_3$ and $v_1$  translates in  hierarchy among the energy mass scale $\mu_3$ and $\mu_1$. For example: if we assume $v_1$ at TeV  and $v_3$ at eV scale, the relation above implies  $\mu_3= 10^{12} \mu_1$ which sounds very weird. Thus,  it seems that the potential above prefers scenarios where both $v_1$ and $v_3$  are not so distant one from another. Since $v_3$ must be small to accommodate standard neutrino masses, then $v_1$ must be small, too. We can conclude that this model  prefers that right-handed neutrinos are  light particles. The most strong reason to the existence of light right-handed neutrinos is the explanation of  short-baseline neutrino results (LSND and MiniBooNE)\cite{miniboone}\cite{lsnd} by means of neutrino oscillation. In this  case, the natural value for $v_1$ is one such that  accommodates at least one right-handed neutrino with mass around eV  with robust mixing with the standard neutrinos and is in conciliation with cosmology. We follow this scenario.  

Such a scenario may be realized for the following set of values for the VEVs,
\begin{equation}
    v_1=10^5\mbox{ eV};\,\,\,\, v_2=246\mbox{ GeV};\,\,\,\,v_3=1\mbox{ eV}, 
    \label{vevvalues}
\end{equation}
and the following set of values for the Yukawa couplings,
\begin{equation}
Y_D =
\begin{pmatrix}
3,6\times10^{-13} &5,74\times10^{-13} & 5,74\times10^{-14} \\
-2,21\times10^{-13} &  -3,45\times10^{-10}  &5,75\times10^{-13} \\
-7,07\times10^{-13}  & 5,75\times10^{-12} & 5,75\times10^{-13}
\end{pmatrix};
\label{YukD}
\end{equation}

\begin{equation}
Y_L =
\begin{pmatrix}
6,20\times10^{-3} &-4,11\times10^{-3} & -1,25\times10^{-2} \\
-4,11\times10^{-3} &  3,90\times10^{-1}  &1,95\times10^{-2} \\
-1,25\times10^{-2}  & 1,95\times10^{-2} & 3,83\times10^{-2}
\end{pmatrix};
\label{YukL}
\end{equation}

\begin{equation}
Y_R =
\begin{pmatrix}
1,40\times10^{-5} &4,75\times10^{-12} & 4,52\times10^{-12} \\
4,75\times10^{-12} & 10^{-1}  &5,08\times10^{-15} \\
4,52\times10^{-12}  & 5,08\times10^{-15} & 10^{-1}
\end{pmatrix}.
\label{YukR}
\end{equation}

On substituting all these values  in $M^{D+M}$, given in Eq. (\ref{M+D}), we have that its diagonalization provides
\begin{eqnarray}
m_{{1}}&&=2\times10^{-4}\;eV;\;\; m_{{2}}=8,6\times10^{-3}\;eV;\;\;
m_{{3}} = 5\times10^{-2}\;eV; \nonumber \\
m_{4}&&=1,4\;eV;\;\;
m_{5}=10^{4}\;eV;\;\;
m_{6}=10^{4}\;eV.
\label{nuprediction}
\end{eqnarray}

The mixing matrix, $U$, responsible by the diagonalization of $M^{D+M}$ and that relates the basis  $\nu$ with $N$, as in Eq. (\ref{mixing}), is given by
\begin{equation}
U = 
\begin{pmatrix}
0,83 &0,54 & -0,12 &0,045 &10^{-5} & 10^{-6}\\
-0,25 &  0,59  &0,72 &-0,03 &-6\times10^{-3} & 10^{-5} \\
0,44 & -0,6 & 0,69 &-0,09 &10^{-4} & 10^{-5} \\
-0,045 &0,03 & 0,09&1 &\sim 0 & \sim 0 \\
10^{-6} &10^{-4} & 10^{-4}& \sim 0 & 1  &\sim 0  \\
10^{-6} &10^{-5} & 10^{-4}& \sim 0  & \sim 0 &1 
\end{pmatrix}.
\label{Mix}
\end{equation}

The values of $m_{{1}}$, $m_{{2}}$ and $m_{{3}}$ given in Eq. (\ref{nuprediction}) and the upper left $3\times3$ submatrix of $U$  accommodate the current solar and atmospheric neutrino oscillations data. A nice thing to observe is that the mixing angles between  $N_4$ , $\nu_{\mu}$ and $\nu_{e}$, together with the mass value of $m_{4}$, are in such a way that they allow the explanation of neutrino anomalies suggested by the data from SBL neutrino experiments by means of neutrino oscillation. Finally, observe  in $U$ that  $N_5$ and $N_6$ practically decouple from the other neutrinos. In other words, this case recovers the $3+1$ sterile neutrino scenario.

A  problem with models involving  eV sterile neutrino is that they present a tension with current cosmological data\cite{cosmotension}. We discuss this point later.

\section{Scalar sector }

	We saw in the previous section that the scenario we are developing here is capable of accommodating neutrino physics including short-baselibe (SBL) anomalies as LSND and MiniBooNE. This provides a strong reason for we go deep into the development of such case. Thus, in this section we perform a careful analysis of the spectrum of the scalars of the model.
\subsection{Spectrum of scalars }
Here we are interested in the spectrum of scalars  for the specific case when $v_3<<v_1<<v_2$. We start developing the  CP-even sector. Considering  the basis $(R_1 \,\,\, R_3 \,\,\, R_2)$, the potential above together with the minimum conditions provide,
\begin{equation}
M_R^2=
\begin{pmatrix}
2\beta_1 v_1^2 + \dfrac{1}{2}\kappa v_2^2 \dfrac{v_3}{v_1} & \beta_3 v_1 v_3 - \dfrac{1}{2}\kappa v_2^2 & \beta_2 v_1 v_2 - \kappa v_2 v_3 \\
\beta_3 v_1 v_3 - \dfrac{1}{2} \kappa v_2 ^2 & 2(\lambda_2 + \lambda_4)v_3^2 + \dfrac{1}{2} \kappa v_2^2 \dfrac{v_1}{v_3} & (\lambda_5 + \lambda_3) v_2 v_3 - \kappa v_1 v_2\\
\beta_2 v_3 v_2 - \kappa v_2 v_1  & (\lambda_5 + \lambda_3) v_2 v_3 - \kappa v_1 v_2 & 2\lambda_1 v_2^2
\end{pmatrix}.
\label{Realmatrix}
\end{equation}  
The complexity of this mass matrix does not allow us to obtain neither the eigenvalues or the eigenvectors. However, according to the hierarchy of the VEVs we assumed here, this matrix may be  approximated by
\begin{equation}
M_R^2 \approx
\begin{pmatrix}
\dfrac{1}{2}\kappa v_2^2 \dfrac{v_3}{v_1} &- \dfrac{1}{2}\kappa v_2^2 & \sim 0 \\
 - \dfrac{1}{2} \kappa v_2 ^2 &  \dfrac{1}{2} \kappa v_2^2 \dfrac{v_1}{v_3} & \sim 0 \\
  \sim 0 & \sim 0  & 2\lambda_1 v_2^2
\end{pmatrix}.
\label{Realmatrixap}
\end{equation}
\\
This means that $R_2$ decouple from the other ones, while $R_1$ and $R_3$ mix among themselves to form  $H_1$ and $H_3$ according to the following relation

\begin{equation}
\begin{pmatrix}
H_1  \\
H_3
\end{pmatrix}
= U_R 
\begin{pmatrix}
R_1  \\
R_3
\end{pmatrix};\;\;
R_2 = H_2,
\end{equation}
where
\begin{equation}
U_R \approx 
\begin{pmatrix}
1 & \epsilon \\
-\epsilon & 1
\end{pmatrix}; \;\; \epsilon \approx \dfrac{v_3}{v_1}.
\label{UR}
\end{equation}

The masses are given by,

\begin{eqnarray}
m_{H_1}^{2}\approx \dfrac{2\beta_2^{2}v_1^{2}}{\kappa},\;\;\;\;\; m_{H_3}^{2} \approx \dfrac{\kappa v_1  v_2^{2}}{2 v_3},\;\;\;\;\; m_{H_2}^{2}\approx 2 \lambda_1 v_2^2.  
\label{massaescalar}
\end{eqnarray}

Observe that, for the hierarchy of the VEVs assumed here, we have that $H_2$ will play the role of the standard Higgs while $H_3$ is a heavy Higgs, with mass around TeV scale, and $H_1$ is a light one with mass at eV scale. 

In the CP-odd sector, things are much simple and the   mass matrix in the basis $(I_1\,\,,\,\, I_2\,\,,\,\, I_3)$ is given by
\begin{equation}
M_I^2 =
\begin{pmatrix}
\dfrac{1}{2}\kappa v_2^2 \dfrac{v_3}{v_1} &\kappa v_2 v_3 & \dfrac{1}{2} \kappa v_2^2 \\
 \kappa v_2 v_3 &  2 \kappa v_1 v_3  &\kappa v_1 v_2 \\
\dfrac{1}{2}\kappa v_2^2   & \kappa v_1 v_2 & \dfrac{1}{2}\kappa v_2^2 \dfrac{v_1}{v_3}
\end{pmatrix}
\label{Immatrix}.
\end{equation}
\\
Its diagonalization leads to a Goldstone boson, $G$, that is dominantly $I_2$ and  will be eaten by the standard gauge boson $Z$; a massless pseudo-scalar, $J$, which we call the Majoron and a heavy pseudo-scalar, $A$, which is dominantly $I_3$. The relation among these pseudo-scalars with the basis is given by  

\begin{equation}
\begin{pmatrix}
J \\
G \\
A 
\end{pmatrix}
= U_I 
\begin{pmatrix}
I_1  \\
I_2  \\
I_3
\end{pmatrix},
\end{equation}
where $U_I$ is given by

\begin{equation}
U_I \approx
\begin{pmatrix}
1 &-2\dfrac{v_3^2}{v_1 v_2} & -\epsilon \\
0 &  1  &-2\dfrac{v_3}{v_2} \\
\epsilon& 2\dfrac{v_3}{v_2} & 1
\end{pmatrix}.
\label{Immix}
\end{equation}
 For the case of interest here, the Majoron is related to the basis in the following way,

\begin{eqnarray}
J \approx I_1 - 2 \dfrac{v_3^2}{v_2 v_1} I_2 - \epsilon I_3,
\end{eqnarray}
which allow we conclude that it is dominantly singlet.

The mass of the pseudo-scalar  $A$ take the following expression, 
\begin{equation}
    m^2_A\approx \dfrac{\kappa v_1  v_2^{2}}{2 v_3},
\end{equation}
which allow we conclude that it is a heavy particle even for the set of VEVs considered here.

In what concern the charged scalars, in considering the basis $(\Delta^+ \,\,\,,\,\,\, \phi^+)$, we have the following mass matrix for these scalars 

\begin{equation}
M_{H^{\pm}}^2 =
\begin{pmatrix}
\kappa v_1 v_3 - \dfrac{1}{2} \lambda_5 v_3^{2} &\dfrac{1}{2\sqrt{2}}v_2(\lambda_5 v_3 - 2\kappa v_1)   \\
 \dfrac{1}{2\sqrt{2}}v_2(\lambda_5 v_3 - 2\kappa v_1) &  \dfrac{1}{4v_3} {v_2}^2(2\kappa v_1 - \lambda_5 v_3) \\
\end{pmatrix}.
\label{ChargedMatrix}
\end{equation}

We can easily diagonalize this matrix and find the physical fields

\begin{equation}
\begin{pmatrix}
G^{\pm} \\
H^{\pm}
\end{pmatrix}
= U_{\pm}
\begin{pmatrix}
\phi^{\pm}  \\
\Delta^{\pm}
\end{pmatrix},
\end{equation}

\begin{equation}
U_{\pm}
\approx 
\begin{pmatrix}
1 &\dfrac{\sqrt{2}v_3}{v_2}   \\
 -\dfrac{\sqrt{2}v_3}{v_2} & 1 \\
\end{pmatrix}.
\end{equation}

We see that there are not any relevant mixing between the charged fields. $G^{\pm}$ is the Goldstone while  $H^{\pm}$ is the simply charged scalar whose mass expression is given by

\begin{equation}
m^2_{H^{\pm}} = \dfrac{1}{4 v_3}(2\kappa v_1 - \lambda_5 v_3)(v_2^2 + 2v_3^2) \approx \dfrac{\kappa v_1  v_2^{2}}{2 v_3}.
\end{equation}
Observe that it must be heavy for the choice of the VEVs used here.  

The doubly-charged scalar acquires the following mass expression
\begin{eqnarray}
m_{\Delta^\pm}^2 = \dfrac{1}{2v_3}(\kappa v_1 v_2^2 - 2 \lambda_4 v_3^2 - \lambda_5 v_2^2 v_3) \approx \dfrac{\kappa v_1 v_2^2}{2 v_3},
\label{m++}
\end{eqnarray}
which must be heavy, too. 

Thus, we see have that, although the VEVs $v_1$ and $v_3$ are much smaller than $v_2$, we have that the scalars that belong to the triplet $\Delta$ are  heavier than the standard-like Higgs and their masses are practically determined by the parameter $\kappa$. This is a consequence of the hierarchy of the VEVs. It is curious that the same hierarchy among the VEVs does the opposite with regard to the scalars belonging to the singlet $\sigma$. The scenario predicts a  light scalar $H_1$. The heavy scalars may be probed at the LHC, while the massless $J$ and light $H_1$ will contribute to the invisible decay channels of  the Higgs and $Z$.

\subsection{Some constraints}

The coupling constants $\kappa$, $\beta_2$, $\lambda_{3,5}$ will play an important role in the RGE-evolution of the quartic coupling of the standard-like Higgs $\lambda_1$. Thus, information on these parameters in the form of constraints is mandatory in order to we  conclude if  the  vacuum of the 3-2-1 model in the regime of low energy scale is stable or not. But before we address this issue, let us investigate the contributions of the light scalars  to the invisible decay of the standard neutral gauge boson $Z$. 

In what  concern the invisible decay of $Z$, the Lagrangian of interest is given by 
\begin{eqnarray}
{\cal L}_{R_3 I_3 Z}\supset	 - \dfrac{g}{c_w} Z^{\mu} [R_3\partial_{\mu}I_3 - I_3\partial_{\mu}R_3].
\end{eqnarray}

Because $R_3$ mix with $R_1$ to compose $H_1$ and $I_3$ mix with $I_1$ to compose $J$, we have that this Lagrangian generates an interaction among $Z$ , $H_1$ and $J$ modulated by the following vertex 
\begin{eqnarray}
V_{ZH_{1}(P_1)J(P_2)}&&\approx \dfrac{g\epsilon^{2}}{c_W}(P_1 - P_2)_\mu ,
\end{eqnarray}
where $g$ is the $SU(2)$ coupling constant and $c_W=\cos(\theta_W)$ with $\theta_W$ being the Weinberg angle. $\epsilon$ is given in Eq. (\ref{UR}). The current data gives $\Gamma(Z)_{inv}=500.1 \pm 1.9 $ MeV \cite{PDG}.  Because $M_{H_1}<<M_Z$, the vertex above provides the following expression for the decay width  $Z\rightarrow H_1 J$,
\begin{eqnarray}
\Gamma(Z\rightarrow J H_1) = \dfrac{  M_Z \epsilon^4 G_F  }{16 \sqrt{2}\pi}(M_Z - \dfrac{M_{H_1}^2}{M_Z})^2 \approx \dfrac{{M_Z}^3 \epsilon^4 G_F  }{16 \sqrt{2}\pi}.
\label{invdec}
\end{eqnarray}

 The expression for the  decay width of $Z$ in two neutrinos is given by

\begin{eqnarray}
\Gamma(Z\rightarrow \bar{\nu} \nu) = \dfrac{ G_F M_{Z}^{3} }{12\sqrt{2}\pi}.
\label{nunu}
\end{eqnarray}

On substituting the current values of the standard parameters that enter in the expression above, i.e.,  $M_Z=91.18$ GeV, $G_F= 1.1663787 \times 10^{-5}$ G$eV^{-2}$ we obtain $\Gamma(Z\rightarrow \bar{\nu} \nu) \approx 166$ MeV. In view of this, the window for new physics is established by $\Gamma(Z)_{inv} - 3\times \Gamma(Z\rightarrow \bar{\nu} \nu) \approx 2.1$ MeV. In other words, all  new contributions to the invisible decay of $Z$ must lie within $2.1$ MeV.

Observe that Eqs. (\ref{invdec}) and (\ref{nunu}) provide

\begin{eqnarray}
\dfrac{\Gamma_{Z\rightarrow JH_1}}{\Gamma_{Z\rightarrow\bar{\nu} \nu}} \approx 0.75 \epsilon^4 \rightarrow \Gamma_{Z\rightarrow JH_1} \approx 124.5 \epsilon^4\,\, \mbox{MeV}.
\end{eqnarray}
 According to this we have that  $\Gamma_{Z\rightarrow JH_1}$ must be smaller than $2.1$ MeV. Once $\dfrac{v_3}{v_1}=\epsilon$, at the end of the day we get
 
 \begin{eqnarray}
 \epsilon<0.36 \rightarrow v_1 > 2.77 v_3.
 \end{eqnarray}
This result confirms the hierarchy among the VEVs we are considering here.

In order to check that our scenario obeys the constraint put by the invisible decay of $Z$ as discussed above, see that for $v_1=10^5$ eV and $v_3=1$ eV, we get  $\Gamma(Z\rightarrow J H_1)=124.5\times10^{-20}$ MeV  which is much smaller than $2.1$ MeV.  The other possible contribution to $\Gamma(Z)_{inv}$ is $\Gamma(Z^{0}\rightarrow JJJ)$. However we must have that $\Gamma(Z^{0}\rightarrow J H_1) > \Gamma(Z\rightarrow JJJ)$ because the later decay is obtained from the first by means of the decay $H_1 \rightarrow JJ$. Thus, we conclude here that the invisible $Z$ decay is not a threat to our model.

Now let us extract constraints over the parameters of the potential by means of the invisible Higgs decay channels and the LFV process $\mu \rightarrow e\gamma$.

Let us consider the contributions that our case give to the invisible decay of the standard-like Higgs $H_2$. We consider the following contributions $\Gamma(H_2\rightarrow H_1H_1)$ and $\Gamma(H_2\rightarrow JJ)$. Their decay  widths  take the expression\cite{valle2016}
\begin{eqnarray}
\Gamma(H_2\rightarrow H_1H_1) \approx \dfrac{\beta_2^{2} v_2}{128\sqrt{2}\pi } \;\;\mbox{and} \;\;\Gamma(H_2\rightarrow JJ) \approx \dfrac{(\lambda_3 + \lambda_5)^{2} v_2}{128\sqrt{2}\pi}.
\end{eqnarray}

The prediction for the  total  decay width of the standard Higgs is around $4$ MeV   with $\sim$20$\%$  being  invisible decay rates(  $BR(H_2 \rightarrow \mbox{inv})=0,26\pm{0,17}$).  All this allows we conclude that $\beta_2$, $\lambda_3$ and $\lambda_5$ are constrained to lie around $10^{-2}$ or smaller.

Thus we conclude here that the 3-2-1 model in the regime of low energy scale, although has a Majoron, which is a massless pseudo-scalar, and a light CP-even scalar it is a safe model in what concern the invisible decay of the standard neutral gauge boson $Z$. As a nice fact we have  that our particular case gives reasonable contribution to the invisible decay of the standard Higgs through the channels  $\Gamma(H_2\rightarrow H_1H_1)$ and $\Gamma(H_2\rightarrow JJ)$. In other words, our case may be constrained by future improvement of the data concerning Higgs physics.

In what concern LFV processes, the muon decay channel $\mu \rightarrow e\gamma$ may provide strong constraints on the parameters of the Potential. In  one-loop order we have the following expression for the branching ratio of this process\cite{BR}

\begin{eqnarray}
BR(\mu\rightarrow \gamma e )\;\;\; \approx\;\;\; \dfrac{ 27\alpha \mid (Y_{L})_{11}(Y_{L})_{12} + (Y_{L})_{13}(Y_{L})_{32} + (Y_{L})_{12}(Y_{L})_{22} \mid^2}{64 \pi  G_F^2 M_{\Delta^{++}}^4},
\end{eqnarray}
where $\alpha$ is the fine structure constant and $G_F= 1.1663787 \times 10^{ -5}$ G$eV^{-2}$ .

On substituting the expression of the mass of  the doubly charged scalar given in Eq. (\ref{m++}), we have that for the fixed  values of $Y_L$'s given in Eq. (\ref{YukL}) and of the  VEVs given in Eq. (\ref{vevvalues}), the upper bound $BR(\mu\rightarrow \gamma e )< 5.7\times10^{-13}$ \cite{MEG}translates in the following lower bound over $\kappa$
\begin{eqnarray}
\dfrac{7\times10^{-19}}{\kappa^{2}}<5.7\times10^{-13}\rightarrow\kappa>1.1\times 10^{-3}.
\end{eqnarray}

With this set of constraints in hand, we are ready to analysis the RGE-evolution of the quartic coupling of the standard-like Higgs $\lambda_1$.

%%%%%%%%%%%%%%
\section{Vacuum Stability }
%%%%
Now that we have developed the scalar sector by finding the spectrum of scalars for a particular set of values of  the VEVs  and obtained some constraints over the parameters of the potential due to Higgs invisible decay and lepton flavor violation, it is the moment to investigate the stability of the vacuum by finding the bound from below conditions and calculating the running of the self coupling of the Higgs.

\subsection{Bound from Below conditions}

In order to assure that the scalar Potential of the 3-2-1 model is bounded from below at large field strength, where the potential 
is generically dominated by the Quartic terms, we need to find the set of conditions that guarantee that the parameters of the Quartic Couplings of the Potential are positive when the fields go to infinity. We find the whole set of conditions and paved the way for similar models.  We follow the techniques employed in \cite{bfbpaper}.
	
Firstly, we separate the quartic couplings of the potential,
   \begin{eqnarray}
	V^{4} = \lambda_1 (\Phi^{\dagger}\Phi)^{2} + \lambda_2 [tr(\Delta^{\dagger} \Delta)]^2 
	+ \lambda_3\Phi^{\dagger}\Phi tr(\Delta^{\dagger} \Delta) + \lambda_4 tr(\Delta^{\dagger} 
	\Delta\Delta^{\dagger} \Delta)  + \lambda_5 (\Phi^{\dagger}\Delta^{\dagger}\Delta\Phi) \nonumber\\
	+ 
	\beta_1 (\sigma^{*}\sigma)^2 + \beta_2\Phi^{\dagger}\Phi\sigma^{*}\sigma + 
	\beta_3 tr(\Delta^{\dagger} \Delta) \sigma^{*}\sigma - \kappa(\Phi^{T}\Delta\Phi\sigma + H.c.),
	\label{equart}
    \end{eqnarray}
and then build the following parametrization:
\begin{eqnarray}
r^2 &&= \Phi^{\dagger}\Phi + tr(\Delta^{\dagger}\Delta) + \sigma^{*}\sigma , \nonumber\\
\Phi^{\dagger}\Phi &&= r^2 cos^2\gamma sin^2\theta ,\nonumber\\
tr(\Delta^{\dagger}\Delta) &&= r^2 sin^2\gamma sin^2\theta ,\nonumber\\
\sigma^{*}\sigma &&= r^2 cos^2\theta,
\end{eqnarray}

where $0 \leq r \leq \infty$, $0\leq \gamma \leq \dfrac{\pi}{2} $ and $0 \leq \theta \leq \dfrac{\pi}{2}$.

We also need to develop the following parameters,

\begin{eqnarray}
\zeta &&= \dfrac{tr(\Delta^{\dagger}\Delta\Delta^{\dagger}\Delta)}{[tr(\Delta^{\dagger}\Delta)]^2} ,\nonumber\\
\xi &&= \dfrac{\Phi^{\dagger}\Delta\Delta^{\dagger}\Phi}{\Phi^{\dagger}\Phi tr(\Delta^{\dagger}\Delta)} ,\nonumber\\
\alpha &&= \dfrac{Re(\Phi^{T}\Delta\Phi\sigma)}{tr(\Delta^{\dagger}\Delta)\sigma^{*}\sigma + \Phi^{\dagger}\Phi \sigma^{*}\sigma + tr(\Delta^{\dagger}\Delta) \Phi^{\dagger}\Phi},
\label{parameters}
\end{eqnarray}
where $ \dfrac{1}{2}\leq \zeta  \leq 1$, $0 \leq \xi  \leq 1$ and $-1 \leq \alpha  \leq 1 $. Two of them are already knew in the literature. The third one is a new parameter. We can see in detail in Appendix A how we can  limit this parameter.

Let also define new  variables $x$ and $y$ that must vary between $0$ and $1$ in the following way:
\begin{eqnarray}
y &&=  sin ^2 \theta ,\nonumber \\
x &&=  sin^2 \gamma. 
\label{eq:1}
\end{eqnarray}

Replacing Eq. (\ref{eq:1}) in Eq. ( \ref{equart}) we get, 

\begin{eqnarray}
\dfrac{V^{4}}{r^4} &&=y^2[\lambda_1 (1-x)^2 + \lambda_2 x^2 + \lambda_3 (1 - x)x + \zeta \lambda_4 x^2 + \xi \lambda_5 (1 - x)x - 2\kappa \alpha x(1-x)] \nonumber \\ 
&&+ (1-y)^2 \beta_1 + (1-y)y [\beta_2 (1-x) + \beta_3 x - 2\kappa \alpha].
\end{eqnarray}

We manage things such that we  can express these quartic terms in the following way,
\begin{eqnarray}
\dfrac{V^{4}}{r^4} &&=A_x y^2+ B_x (1-y)^2  + C_x (1-y)y \,\,\,\,\, \mbox{where},\nonumber \\
A_x &&= \lambda_1 (1-x)^2 + (\lambda_2  +\zeta \lambda_4)x^2 + (\lambda_3  + \xi \lambda_5  - 2\kappa \alpha)(1 - x)x  ,\nonumber \\
B_x &&= \beta_1  ,\nonumber \\
C_x &&=  \beta_2 (1-x) + \beta_3 x - 2\kappa \alpha.
\label{eqconditions}
\end{eqnarray}

We can fix $y=0$ or $y=1$ to obtain the cases when the quartic couplings of the potential is positive. When we do this, we obtain the following conditions

\begin{eqnarray}
 A_x > 0,
 \label{a}
 \end{eqnarray}
\begin{eqnarray}
 B_x > 0 ,
 \label{b}
\end{eqnarray}
 \begin{eqnarray}
 C_x + 2 \sqrt{A_x B_x} > 0.
\label{c}
\end{eqnarray}

For $A_x > 0$ we need to use the same argument as before. Fixing $x=0$ and $x=1$ we have similar conditions for the inequalities
\begin{eqnarray}
&&\lambda_1 > 0 ,\nonumber \\
&&\lambda_2 + \zeta \lambda_4 > 0 ,\nonumber\\
&&\lambda_3 + \xi \lambda_5 - 2 \kappa \alpha + 2 \sqrt{\lambda_1(\lambda_2 + \zeta \lambda_4)} > 0. 
\end{eqnarray}

These new conditions depends of the  parameters in Eq. (\ref{eqconditions}). They vary in different ranges, but we only need to study the boundary values of these intervals. In this case the new conditions are:

\begin{eqnarray}
&&\lambda_1 > 0 ,\nonumber \\
&&\lambda_2 + \lambda_4 > 0 ,\nonumber\\
&&\lambda_2 + \dfrac{1}{2}\lambda_4 > 0 ,\nonumber\\
&&\lambda_3 + 2 \kappa  + 2 \sqrt{\lambda_1(\lambda_2 + \dfrac{1}{2} \lambda_4)} > 0 ,\nonumber\\
&&\lambda_3 + 2 \kappa  + 2 \sqrt{\lambda_1(\lambda_2 + \lambda_4)} > 0 ,\nonumber\\
&&\lambda_3 + \lambda_5 + 2 \kappa + 2 \sqrt{\lambda_1(\lambda_2 + \dfrac{1}{2} \lambda_4)} > 0 ,\nonumber\\
&&\lambda_3 + \lambda_5 + 2 \kappa + 2 \sqrt{\lambda_1(\lambda_2 +  \lambda_4)} > 0 ,\nonumber\\
&&\lambda_3 - 2 \kappa  + 2 \sqrt{\lambda_1(\lambda_2 + \dfrac{1}{2} \lambda_4)} > 0 ,\nonumber\\
&&\lambda_3 - 2 \kappa  + 2 \sqrt{\lambda_1(\lambda_2 + \lambda_4)} > 0 ,\nonumber\\
&&\lambda_3 + \lambda_5 - 2 \kappa + 2 \sqrt{\lambda_1(\lambda_2 + \dfrac{1}{2} \lambda_4)} > 0 ,\nonumber\\
&&\lambda_3 + \lambda_5 - 2 \kappa + 2 \sqrt{\lambda_1(\lambda_2 + \lambda_4)} > 0 .
\end{eqnarray}

Using the same argument for the condition in Eq. ( \ref{b}), it turns easy to see that
\begin{eqnarray}
 \beta_1 > 0. 
\end{eqnarray}

Using the condition in Eq. (\ref{c}) and the same fact that $C_x$ can have $x=0$ or $x=1$, we obtain
\begin{eqnarray}
&&\beta_2 - 2\kappa\alpha + 2 \sqrt{\beta_1 \lambda_1} > 0 ,\nonumber \\ 
&&\beta_3 - 2\kappa\alpha + 2 \sqrt{\beta_1 (\lambda_2 + \zeta \lambda_4))} > 0.
\end{eqnarray}

The first inequality has two different solutions while the second has four ones.  At the end of the day, we have
\begin{eqnarray}
&&\beta_2 + 2\kappa + 2 \sqrt{\beta_1 \lambda_1} > 0 ,\nonumber \\
&&\beta_2 - 2\kappa + 2 \sqrt{\beta_1 \lambda_1} > 0 ,\nonumber \\
&&\beta_3 + 2\kappa + 2 \sqrt{\beta_1 (\lambda_2 + \dfrac{1}{2} \lambda_4)} > 0 ,\nonumber \\
&&\beta_3 + 2\kappa + 2 \sqrt{\beta_1 (\lambda_2 +  \lambda_4)} > 0 ,\nonumber \\
&&\beta_3 - 2\kappa + 2 \sqrt{\beta_1 (\lambda_2 + \dfrac{1}{2} \lambda_4)} > 0 ,\nonumber \\
&&\beta_3 - 2\kappa + 2 \sqrt{\beta_1 (\lambda_2 +  \lambda_4)} > 0. 
\end{eqnarray}
So, those are the set of condition that guarantee the potential in Eq. (\ref{potential}) is bounded from below. In what follow we obtain the  running of the self coupling related of the standard-like Higgs.

\subsection{RGE-evolution of the self coupling of the standard-like Higgs}

The standard model  predicts that the self coupling  of the Higgs becomes negative  at an energy scale  around $\Lambda =10^{11}$GeV.  This means that the standard model can not assure the stability of the vacuum up to the Planck scale. This must be remedied  by means of new physics in the form of new particles with appropriate interactions. This issue has been extensively investigated in the literature\cite{okada}. Within our scenario we show that the right behavior  of the self coupling of the Higgs that guarantees stability of the Electroweak Vacuum up to Planck scale depends strongly  on the coupling  $\kappa$. We do our analysis by  implementing the model in SARAH 4.13.0 \cite{sarah} and evaluating the $\beta$ function for $\lambda_1$ at one-loop level.

The main contributions for the beta function of $\lambda_1$ involve the following  terms
\begin{eqnarray}
\beta_{\lambda_1} &= & \dfrac{27 }{100}g_Y^4 + \dfrac{9 }{4}g^4  + 
 \dfrac{9}{10} g_Y^2 (g^2 - 2 \lambda_1) - 9 g^2 \lambda_1 + 12 \lambda_1^2 + 12\lambda_1 y_t^\dagger y_t -
 12 y_t^\dagger y_t y_t^\dagger y_t +\nonumber \\
 &&+ 2 \beta_2^2 + 6 \lambda_3^2 - 4 \lambda_3 \lambda_5
 + 2 \lambda_5^2 + 4 \kappa^2 ,
\end{eqnarray}
where $g$ and $g_Y$ are the gauge couplings of the standard gauge group $SU(2)$ and $U(1)_Y$ while  $y_t$ is the Yukawa coupling of the quark top.

\begin{figure}
\includegraphics[scale=0.5]{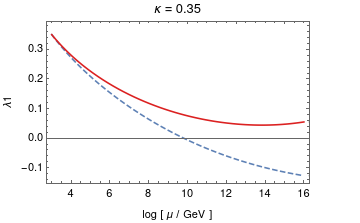}
    \includegraphics[scale=0.5]{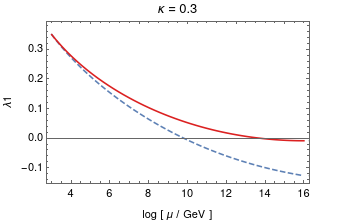}
    \caption{Running of $\lambda_1$ at one-loop level as a function of the energy scale $\mu$ for $\lambda_5=\lambda_3=\beta_2$ =0.001 with $y_t=0.9965$, $g_Y=0.4627$ and $g=0.6535$. The doted line represents the expectation of Standard Model and the red line represents the expectation for our model for two values of $\kappa$.}
    \label{rge1}
\end{figure}

Observe that the couplings $\beta_2$, $\lambda_{3,5}$ and $\kappa$  give  positive contributions to the running of $\lambda_1$. However, as showed above,  the invisible Higgs decay  requires $\beta_2$ , $\lambda_{3,5}$ be minor then  $10^{-2}$ which turns insignificant they contributions to the RGE-evolution. Rest us the contribution of the parameter $\kappa$. In Fig. 1 we show the plot of the running of $\lambda_1$ with energy scale for two possible values of $\kappa$. We see that the running of $\lambda_1$ may get positive up to Planck scale for $\kappa>0.3$.   Thus, the model may have the vacuum stable up to Planck scale thanks to the contribution of the parameter $\kappa$.
%%%%%%%%%%
\section{Concluding remarks}

In this work we studied  stability of the vacuum in the 3-2-1 model with right-handed neutrinos. Our investigation was restricted to a specific scenario characterized  by spontaneous violation of the lepton number at low energy scale. The case is well motivated because it yields light sterile neutrinos and may   explain MiniBooNE  by means of neutrino oscillation.  In such a scenario, we obtained the whole set of conditions that guarantee the model is Bounded From Below and studied the RGE-evolution of the self-coupling of the standard-like Higgs. As main result we have that  the quartic coupling $\kappa \Phi^T \Delta \Phi \sigma$ plays a central role in the process and stability of the  vacuum requires  $\kappa > 0.3$.

As interesting consequence,  we remark that the model has one  Majoron ($J$) and one  light Higgs ($H_1$) composing the spectrum of  scalar of the model. Their  contributions to the invisible  decay rate of the standard-like Higgs,   $H_2 \rightarrow JJ$ and $H_2 \rightarrow H_1 H_1$, were considered and the results are the bounds $\beta_2 , \lambda_{3},\lambda_5 \leq 10^{-2} $ over the couplings of the potential.

In what concern the neutrino sector, the scenario recovers the 3+1 sterile neutrino model which explain MiniBooNE experiment by means of neutrino oscillation.  However, we know that light sterile neutrinos are strongly disfavored by current cosmological data involving Big Bang Nucleosynthesis(BBN) , Cosmic Microwave Background(CMB) anisotropies and Large Scale Structure(LSS)\cite{cosmotension}. This is so because, in face of the large mixing required by MiniBooNE, neutrino oscillation may conduct sterile neutrino to thermal equilibrium with the active neutrino even before neutrinos decouple from the primordial plasma. A possible solution for this tension requires the suppression of the production of these neutrinos in the early universe. This avoids that they thermalize with the active ones at high temperature. This may be achieved by means of secret interactions\cite{SI} which is nothing more than the interaction of the sterile neutrino with a pseudo-scalar, $I$,
\begin{equation}
    \sim g_s \bar \nu^C_S\gamma_5 \nu_S I. 
    \label{secretinteracvtion}
    \end{equation}
 The solution to the tension requires $I$ be lighter than the lightest sterile neutrino and  $g_s$ take values in the range  $10^{-6}-10^{-5}$.  Observe that our scenario recover this solution. For this, recognize that $g$ is $Y^R_{11}$ whose value in the matrix in Eq. (\ref{YukR}) is $1,4\times 10^{-5}$ and $I$ is the Majoron $J$. In order to generate a small mass to $J$ we just need to consider a term like:  $M \sigma \sigma \sigma$ in the potential. This term will generate a mass term to $J$ proportional to $M$. On assuming that $M<  m_{N_4}$ we have a secret sector that reconciliates eV sterile neutrino with cosmology as done in \cite{piresSI}.

\section{Appendix A}
Here we will give a hint for the proof of the limitation of the parameter $\alpha$. The definition of this parameter is
\begin{eqnarray}
&&\alpha  = \dfrac{Re[\Phi^{T}\Delta \Phi \sigma]}{\Phi^{\dagger}\Phi \sigma^{\dagger}\sigma + tr[\Delta^{\dagger}\Delta]\Phi^{\dagger}\Phi  + tr[\Delta^{\dagger}\Delta]\sigma^{\dagger}\sigma}.
\end{eqnarray}

We can expand this parameter in terms of the components of the fields. We have that

\begin{eqnarray}
\mbox{Numerator} &&= Re[\phi^{0}\Delta^{0} \phi^{0} \sigma + \sqrt{2}\phi^{0}\Delta^{+} \phi^{-} \sigma + \phi^{-}\Delta^{++} \phi^{-} \sigma ],\nonumber\\  
\mbox{Denominator} &&= (\phi^{0\dagger}\phi^{0} + \phi^{+}\phi^{-} +\sigma^{\dagger}\sigma)( \Delta^{0\dagger}\Delta^0 + \Delta^+\Delta^- + \Delta^{++}\Delta^{--} )\nonumber \\
&&+ \sigma^{\dagger}\sigma(\phi^{0\dagger}\phi^{0} + \phi^{+}\phi^{-}). 
\end{eqnarray}

Then, we can study term by term to see what is the behavior of this parameter, e.g., to see if it is limited or not. As an example, we choose the first term of the Numerator and expand the fields in the real and imaginary parts. Using the following expansion
\begin{eqnarray}
\phi^0 &&= R_2 + i I_2 ,\nonumber \\
\Delta^0 &&= R_3 + i I_3 ,\nonumber \\
\sigma &&= R_1 + i I_1 ,
\end{eqnarray}
we will obtain the Denominator terms (only the real part)
\begin{eqnarray}
&&R_2^2R_3R_1 - R_2R_3I_2I_1 - I_2I_3R_2R_1 + I_1 I_2^2 I_3 - I_2^2 R_1R_3 
- I_1I_2 R_2 R_3 - R_1 R_2R_2I_3 - R_2^2 I_1I_3.
\nonumber
\end{eqnarray}

The idea here is to look closely in each real function and study their limitation range. For the first term, $R_2^2R_3R_1$, we have the following relation (for $R_2 \neq 0$ )

\begin{eqnarray}
\dfrac{R_2^2R_3R_1}{R_2^2 R_3^2 + R_1^2 R_2^2 + R_1^2 R_3^2 + (...)} \rightarrow \dfrac{R_3R_1}{R_3^2 + R_1^2 + \dfrac{R_1^2 R_3^2}{R_2^2} + (...)} < \dfrac{R_3R_1}{R_3^2 + R_1^2}.
\end{eqnarray}

We can see easily that this last term is limited in the range [-1,1] with polar coordinates. We use similar arguments for next terms and find that $\alpha$ lies in the range [-1,1].

\acknowledgments
J.P.P thanks CNPq for financial support. C.A.S.P  was supported by the CNPq research grants No. 304423/2017-3. 

%%%%%%%%%%%%%%%%%%%%%%%%%%%%%%%%%%%%%%%%%%%%%%

%%%%%%%%%%%%%%%%%%%%%%%%%%%%%%%%%%%%%%%%%%%%%%%%%%%%%%%%%%%%%%%%%%%%%%%%%%%%%%%%%%%%%%%%%%


\begin{thebibliography}{99}
%%%%%%%
\bibitem{2hdm}
G. C. Branco , P. M. Ferreira, L. Lavoura, M.N. Rebelo, Marc Sher, Joao P. Silva, Phys. Rept.{\bf 516} (2012) 1-102.
%%%%%%%%%
\bibitem{cheng}
T. P. Cheng, Ling-Fong Li, Phys.Rev.Lett. {\bf 45} (1980) 1908;
G. B. Gelmini, M. Roncadelli, Phys. Lett. B{\bf 99} (1981) 411-415.
%%%%%
\bibitem{singlet}
N. V. Krasnikov, Phys. Lett. B{\bf 291}(1992) 89;
A. S. Joshipura and J. W. F.Valle, Nucl. Phys; B{\bf 397}(1993) 105;
Donal O'Connell, Michael J. Ramsey-Musolf, Mark B. Wise, Phys.Rev. D{\bf 75} (2007) 037701.
%%%%%%
\bibitem{Schechter}
J. Schechter and J. W. E Valle, Phys. Rev. D {\bf 25} (1982) 774.
%%%%
\bibitem{valle}
M. A. Diaz, M.A. Garcia-Jareno, Diego A. Restrepo, J.W.F. Valle, Nucl. Phys. B{\bf 527} (1998) 44-60;
Cesar Bonilla, Jorge C. Romão, José W. F. Valle, New J. Phys.{\bf 18} (2016) no.3, 033033;
Sylvain Blunier, Giovanna Cottin, Marco Aurelio Díaz, Benjamin Koch, Phys. Rev. D{\bf 95} (2017) 075038.
%%%%
\bibitem{cogollo}
For an extension of this scenario involving 2HDM model, see:
D. Cogollo, Ricardo D. Matheus, T\'essio B. de Melo, Farinaldo S. Queiroz,  Phys. Lett. B{\bf 797} (2019) 134813.
%%%%
\bibitem{miniboone}
A. Aguilar-Arevalo, {\it et al.}, Phys. Rev. Lett. {\bf 121} (2018) 221801. 
%%%
\bibitem{lsnd}
A. Aguilar-Arevalo, {\it et al.}, Phys. Rev. D{\bf 64} (2001) 112007.

\bibitem{cosmotension}
 G. Steigman, Adv. High Energy Phys.{\bf 2012} (2012) 268321;
N. Aghanim, (Planck Collaboration), {\it et al.}, arXiv:1807.06209;
 J. Hamann, S. Hannestad, G. G. Raffelt, Y. Y.Y. Wong, JCAP{\bf 1109} (2011) 034. 
%%%%
\bibitem{PDG}
M. Tanabashi {\it et al.} (Particle Data Group), Phys. Rev. D {\bf 98}(2018) 030001  and 2019 update.
%%%
\bibitem{BR}
A. G. Akeroyd, Mayumi Aoki, and Hiroaki Sugiyama, Phys. Rev. D {\bf 79} (2009) 113010;
For a general formulae for $f_1\rightarrow f_2 \gamma$, see: L. Lavoura, Eur. Phys. J. C {\bf 29} (2003) 191.




\bibitem{valle2016}
Cesar Bonilla, Jorge C. Romão, José W.F. Valle,  New J.Phys. 18 (2016) no.3, 033033 hep-ph/1511.07351.


\bibitem{valle1998}
 Marco A. Diaz, M. A. Garcia-Jareno, Diego A. Restrepo, J.W.F. Valle, Nucl. Phys. B{\bf527}:44-60 (1998) hep-ph/9803362.

\bibitem{invdecay}
M. Aaboud et al. (ATLAS Collaboration)
Phys. Rev. Lett. {\bf122}, 231801. 
%%%%
\bibitem{MEG}
MEG Collaboration, Phys. Rev. Lett. {\bf 110} (2013) 201801.

\bibitem{bfbpaper}
A. Arhrib, R. Benbrik , M. El Kacimi, L. Rahili, S. Semlali, Eur. Phys.J. C{\bf 80} (2020) 13;
Daniel A. Camargo, Alex G. Dias, Tessio B. de Melo, Farinaldo S. Queiroz, 
hep-ph/1811.05488.

\bibitem{okada}
Naoyuki Haba, Hiroyuki Ishida, Nobuchika Okada, Yuya Yamaguchi,  Eur. Phys. J. C  {\bf76} (2016) 333;
Cesar Bonilla, Renato M. Fonseca, José W. F. Valle, Phys. Rev. D{\bf 92} (2015) 075028.
%%%
\bibitem{sarah}
 F. Staub, Comput. Phys. Commun.{\bf 185}, 1773 (2014) [arXiv:1309.7223 [hep-ph]].
%%%%%
\bibitem{SI}
S. Hannestad, R. S. Hansen, and T. Tram,
Phys. Rev. Lett. {\bf 112}(2014) 031802;
Maria Archidiacono {\it et al.}, JCAP{\bf08}(2016)067;
B. Dasgupta and J. Kopp,
Phys. Rev. Lett.{\bf 112}(2014) 031803;
Xiaoyong Chu {\it et al.}, JCAP{\bf 11}(2018) 049. 
%%%%%
\bibitem{piresSI}
C. A. de S. Pires, Phys. Lett. B{\bf 800} (2020) 135135.




\end{thebibliography}
\end{document}